\documentclass{article}

\usepackage{amsmath}
\usepackage{subcaption}
\usepackage{caption}
\usepackage{amsfonts}
\usepackage{CJKutf8} %
\usepackage[usenames, dvipsnames]{color}
\usepackage{booktabs}
\usepackage{algorithm2e}
\usepackage{fancyhdr}
\usepackage{graphicx}
\usepackage{hyperref}
\usepackage{multirow}

\usepackage{skak} %
\usepackage{tabularx}
\usepackage[thinlines]{easytable} %
\usepackage{tikz}
\usepackage{times}
\usepackage[shortlabels]{enumitem}
\usepackage{placeins}
\usepackage[title]{appendix}
\usetikzlibrary{arrows.meta}

\definecolor{darkgreen}{RGB}{0, 125, 0}
\definecolor{orange}{RGB}{255, 125, 125}

\definecolor{colorwin}{RGB}{156,255,161}
\definecolor{colordraw}{RGB}{220,220,255}
\definecolor{colorloss}{RGB}{255,161,156}

\title{How to cluster nearest unique nodes from different classes using JJCluster in Wisp application?}

\author{
	\href{https://jimut123.github.io/}{Jimut Bahan Pal}\\
	\normalsize{jimutbahanpal@yahoo.com}\\
	\normalsize{Department of Computer Science}\\
	\normalsize{Ramakrishna Mission Vivekananda Educational and Research Institute}
	\\
	\normalsize{Howrah 711202} \\
	\normalsize{ Alumni of \href{https://www.sxccal.edu/SXC-Research/sxc-Publications.htm}{St. Xavier's College}, Kolkata.} \\
}

\date{}

\begin{document}

\maketitle

\begin{abstract}
	The work of finding the best place according to user preference is tedious task. It needs manual research and lot of intuitive processes to find the best location according to some earlier knowledge about the place. It is mainly about accessing publicly available spatial data, applying a simple algorithm to summarise the data according to given preferences, and visualising the result on a map. We introduced JJCluster to eliminate the rigorous way of researching about a place and visualizing the location in real time. This algorithm successfully finds the heart of a city when used in Wisp application. The main purpose of designing Wisp is to find the best location according to a set of preferences in a map. This has interesting application in finding the perfect location for a trip to unknown place which is nearest to a set of preferences. We also discussed the various optimization algorithms that are pioneer of today’s dynamic programming and the need for visualization to find patterns when the data is cluttered. Yet, this general cluster algorithm can be used in other areas where we can explore every possible preference to maximize its utility.
\end{abstract}


\section{Introduction}

\subsection{Background}

To visit an unknown important place we search hotel which is close to a set of preferences and we plan accordingly to explore too many places in a short time. The preferences may be river, museum, shopping mall, lake, cafe, hospital or even movie theatre. There are many possibilities of selecting hotel, but we want the hotel which is only adjoining to every preference. We need the result fast without manually researching about the place. Here, we consider our preferences as a class. Intuition would tell us to search in Google maps to select restaurant. The approach may not be perfect always, and will take a lot of time. Here the main aim of designing Wisp \cite{R1} ~\cite{R2} application is to visualise the place according to user preferences in real time, without any botheration of researching manually. We introduced here an algorithm for finding the cluster (the boundary location), named as JJCluster (which stands for Jimut-Jisnoo Cluster). The software module takes a list of places containing corresponding class, name, latitude, longitude, and displays a map for the Cluster formed at real time. This algorithm can be used in a variety of research areas of general clustering, where we need to select unique closest nodes from each class.

The JJCluster marks a boundary across the location which will be the best place for those preferences. Selecting a coordinate inside that boundary may be one of the best places for stay. The other applications of this software may find developed areas according to certain set of preferences. Major factors that boosted the development of this Wisp application are discussed below.

\subsection{Availability of data}

Cheap storage ~\cite{R3} and sensors have made it possible to collect variety of data in huge quantities. Users generate a large number of data in social networking platforms. The Wisp application used Marker Cluster as its core algorithm until the design of JJCluster.  Beilschmidt et al. (2017) ~\cite{R4} proposed Marker Cluster algorithm to cluster huge dataset into a Map as circles. The quantity of data points may lead to visual clutter and most of the browser may not support such data points in a map, which would result in occlusion and loss of information. The shape of circle in Marker Cluster changes dynamically on zooming in or out of the map, which helped in optimising the available resources and visualising any number of points. By using Marker cluster we couldn’t determine whether a place have all of the preference classes or some of them are in large quantity, so it may lead to wrong visualisation and selection of hotspots.

Similarly, location based social networks (LBSN) have recently gained popularity to help targeted users ~\cite{R5} to get their needs. The use of local experts, the natives, to review and suggest areas of interest to users with similar behavioural pattern has been the most advanced way of recommending places. Semantic similarities and geographic correlations between users are captured simultaneously by the concept of local experts. This technique is used widely for collecting data for application programming interfaces (API’s) which helps in designing of Wisp application.

The following section discusses about the related work done by researchers in various fields of data visualisation (both in 2D and 3D), dynamic programming for resource allocations, and designing of recommendation systems to targeted users. 

\section{Related Works}

Visualizations can help human to predict patterns ~\cite{R6} and areas of concern in a dataset, which may have inherently hidden relations. It gives ideas for processing the data further. It may help policy makers ~\cite{R7}, volunteers, programme managers, donors and non-governmental organisations to invest money in critical places by getting information from visualizations. In the domain of big data, scientists use statistical and data mining techniques ~\cite{R3} to find the information in which they are interested. When the data is huge scientists use certain partitioning techniques to visualise the data along with clustering methods. Big data analysis provides opportunity to test hypothesis with new data, which have potential to yield new insights to classical research questions across multiple disciplines ~\cite{R8}. Old methods such as surveys are being replaced by sensor data, which needs a lot of pre-processing to find valuable and accurate information that was not available by using old methods. There was constant breach of privacy, when crowd sourced data was collected. Quality measures and pre-processing ~\cite{R9} can help to minimise such privacy risks. This includes basic steps of formatting, cleaning, validation, making the data anonymous before publishing. Recent years have boosted the popularity of location based services, where users generate huge amount of data through social media. Such data can be used for recommending important places ~\cite{R5} for class of users. Geodemographic ~\cite{R10} may help to find social similarities amongst variables within a dataset using cluster analysis and profile matching, which are indicators of social, economic and demographic conditions of neighbourhood. These forms the basis of building crowd sourced platforms for data collection which are necessary for visualising data in large scale.

Researchers built a cyber-infrastructure ~\cite{R11} solution named PolarGlobe, which enables comprehensive collaboration and analysis. Their solution does not require installation of plug-ins, since that is a web based framework. State of the art tools used for measurement of climate change risk, adaptation ~\cite{R12} and vulnerability assessment use web as their common infrastructure. The data driven approach analysis is dominating in research of transportation; with the increase in amounts of data collected from accurate transport system sensors, which was earlier determined with the help of mathematical equations due to scarcity of data. In this way data are collected for gaining insights to improve existing systems.

\subsection{Dynamic programming algorithms}

Dynamic programming algorithms are useful in different fields of study ranging from telecommunications, transportations, electrical power systems, water supply, and emergency services which provide useful human resources to certain areas. In designing the optimisation route of network flow there are some constraints, such as the topological characteristics of a networked infrastructure where we choose the best path that is feasible for locomotion. Computer scientists, economist, statisticians, cartographic researchers made significant contributions for finding the optimal path and its solutions for a given set of problem. Optimisations can even help in designing cities by putting areas of critical need such as hospitals ~\cite{R13} in those places which are accessible very quickly. Even allocation of charging stations for electric vehicles ~\cite{R14} are promising solution to tackle climate change impacts for improving air quality and enhancing growth sustainability. Management and location of underground ~\cite{R15} such as depth of cables and pipelines are great concern in many countries, misinformation of which can even result in tragic accidents.

Bellman was one of the earliest scientists who introduced dynamic programming ~\cite{R16} ~\cite{R17}. He worked on dynamic programming problem for finding the shortest path between two points in a graph of interconnected roads and varied density of traffic. This marked the beginning of a new field of optimisation problem known as the dynamic programming problem. Scientists subsequently worked on the problem of optimising certain features in a problem. The most famous problem is Travelling Salesman Problem or the TSP, which is still an extremely interesting topic about whether there exists any algorithm for which can we do better? There is Floyd-Warshall algorithm for finding the shortest path between any two paths in a graph. This algorithm can help to find the distance between any two points of interest in a graph. Scientists can use optimisation when there are interdiction scenarios in a network.

\subsection{Cartographic visualizations for pattern recognition}

Dodge, Weibel, and ~\cite{R18} used two phased approach for studying movement pattern in more systematic and comprehensive way. They developed a conceptual framework for movement behaviour of different moving objects, and then classifying it. They devised a toolbox for the study of data mining algorithms for the movement of animal, human, traffic, industry goods, bee etc. which follow a generic spatiotemporal pattern.

Delort ~\cite{R19} examined a technique for visualising large spatial dataset in Web Mapping Systems. The algorithm created a cluster tree, which helped to visualising the cluster in a map without cluttering itself, at any given scale. Though, the algorithm was too slow for its time, it was later improved by many researchers and used widely till date, this is the basis of the idea for designing JJCluster (coincidentally), but here we focus on quality (classes) over quantity. Phan, Xiao, Yeh, Hanrahan, and Winograd ~\cite{R20} made an algorithm using binary hierarchical clustering to formulate flow maps which are used immensely in studying and visualising migration of birds, people, flow of traffic in a network, flow of river etc. Moreover, Stachoň, Šašinka, Čeněk, Angsüsser, Kubíček, Štěrba, and Bilíková ~\cite{R21} also studied the selection of graphic variables, in a new perspective to use graphics in cartographic visualisations. We used circle markers in the visualisation of this application, along with dynamic legend generators.

Zhu, and Xu ~\cite{R22} reduced the influence of location ambiguity using geotags and geo-cluster to employ semantic features of geo-tagged photographs through clustering. This method measures the semantic correlation which has synonym representation of tags, but not same tags. Delmelle, Zhu, Tang, and Casas ~\cite{R23} proposed a web based open source geospatial toolkit, which are particularly needed in developing countries lacking GIS software. It provides a user friendly and interactive web based framework for analysing the outbreak of vector borne diseases. They analysed the outbreak of dengue for efficient and appropriate allocation of human resources to minimise them. Etten and Jacob ~\cite{R24} et al. provided a package in R for calculating routes and distances in calculating various heterogeneous geographic spaces represented as grids. They used circuit theory and geographical analysis to find out various distances between routes.

\subsection{Recommendation Systems to targeted users}

D’Ulizia, Ferri, Formica, and Grifoni ~\cite{R25} enabled the relaxation of three kinds of query constraints, namely semantic, structural and topological, by using GeoPQL which translated natural language queries to map visualisations. They used graph theoretic procedures for query approximations. It helped those users who were unaware of exactly what they were looking for due to complexity of finding and locating the data points with naive tags. Cai, Lee, and Lee ~\cite{R26} studied the movement of humans to get better semantic knowledge about a place, through the help of GPS and social media. Dareshiri, Farnaghi, and Sahelgozin ~\cite{R27} improved the existing recommendation capabilities of geo portals. In their approach, the geoportals analyses the user’s activity and recommend geospatial resources to users, based on their desires and preferences. Zhao, Qin, Ye, Wang, and Chen ~\cite{R28} identified the centre of cluster in map with the real time traffic data of taxi. These data were further used to study the popularity importance of a place over certain duration of time. The study showed the crowded areas, which were needed by the urban transportation and planning management authorities to reduce congestions.

\subsection{Visualising geospatial data in 3D}

Visualisation in 3D helps to find information, which is almost impossible to find by looking at the data. Three dimensional representations of Geographic information on computers are known as GeoVES ~\cite{R29} or Geospatial Virtual Environment. These environments can engage participants to explore virtual landscapes and knowledge-scapes along with providing different images of reality. Azri, Anton, Ujang, Mioc, and Rahman ~\cite{R30} studied various issues that were needed to be resolved for using geospatial data infrastructure. They optimised space, time accession of data, data model, and spatial access method etc., for utilising and managing high precision laser data, which could take 63 GB of space in a hard drive. The analysis of 3D visual data with the help of existing 3D software tools such as Autodesk Maya can help to find the related pattern in the data which can be used carefully by researchers to provide solutions to existing problems. The approach contributes to intuitive comprehension of complex geospatial information for decision makers in security agencies as well as for authorities related to urban planning for certain correlation between the urban features.

\section{Materials and Methods}

\subsection{Working of the algorithm}

\begin{figure}
	\begin{subfigure}{.5\textwidth}
		\centering
		\includegraphics[totalheight=6cm]{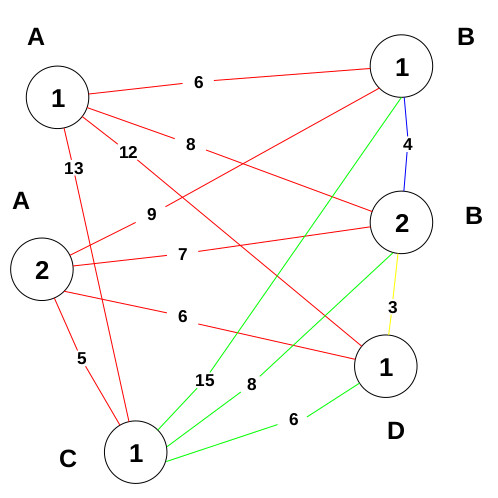}
		\caption{The original graph in the beginning.}
		\label{fig1}
	\end{subfigure}%
	\begin{subfigure}{.5\textwidth}
		\centering
		\includegraphics[totalheight=6cm]{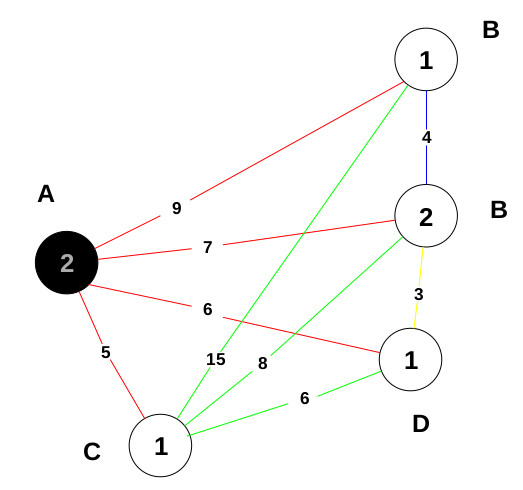}
		\caption{The nearest point from the first class is selected.}
		\label{fig2}
	\end{subfigure}
	
	\begin{subfigure}{.5\textwidth}
		\centering
		\includegraphics[totalheight=4cm]{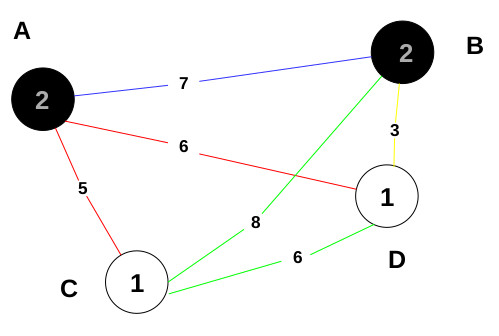}
		\caption{The closest point from the second class to the first class is selected.}
		\label{fig3}
	\end{subfigure}%
	\begin{subfigure}{.5\textwidth}
		\centering
		\includegraphics[totalheight=4cm]{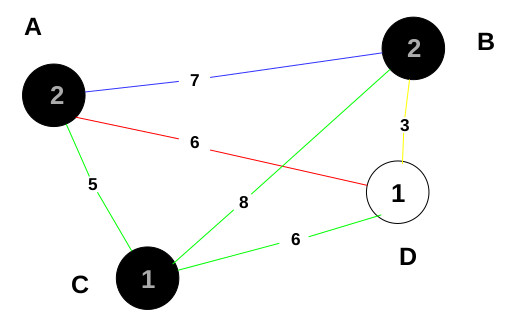}
		\caption{The third point (which is closest since it is only present once) from the third class to the first and second chosen points of their classes is selected.}
		\label{fig4}
	\end{subfigure}
	\caption{A figure containing the structure of graphs formed by selecting the nodes from different classes.}
	\label{fig:nodes}
\end{figure}

\begin{figure}
	\centering
	\includegraphics[totalheight=6.5cm]{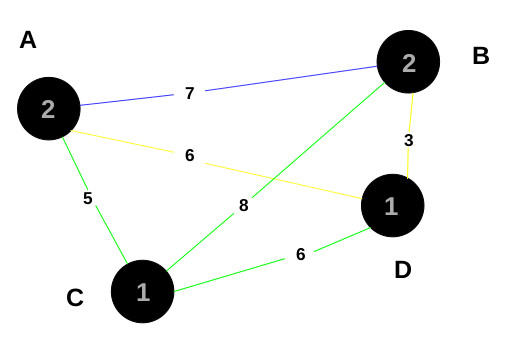}
	\caption{The fourth point from the fourth class (which is closest since it is only present once) from all the selected classes is chosen.}
	\label{fig5}
\end{figure}

The algorithm is very simple, once we write it with a pen and paper. We consider each point in the map as node of a graph. The aim is to select the shortest distance from each of the individual classes. We describe the general working of the algorithm in Fig. ~\ref{fig1},~\ref{fig2},~\ref{fig3},~\ref{fig4} and ~\ref{fig5}. Firstly, we start by choosing the elements of the first class in the graph as shown in Fig. ~\ref{fig1} (the list S is coloured as black in the graph for every step). We choose element 1 from class A and calculate its distance from all the items of the other classes, i.e., from 1 to B’s 1, B’s 2, C and class D. Then we sum up, i.e., 6+8+12+13 = 39 and store it in a list. We do the same thing for node 2 of class A; we get the distance 9+7+6+5 = 27 and add it to the list. Since, there is no other element from the first class, we will choose the minimum from this list i.e., class A’s point 2 as shown in Fig. ~\ref{fig2}. We store this in the list of selected nodes, called S. We can now say that this point in S is nearest to all the other points of every other class. So, the other points of other classes will be nearest to this point. Again, we calculate the same thing for class B’s node 1, but here from point 2 of class A (which is the only element present in the list S), we store it in an empty list. The value that we got is 9. We do this for the second element of class B, and store 7 to the list. Now, we select the nearest point from the entire list S, i.e., node 2 from class B. We store this to the list S as shown in Fig. ~\ref{fig3}. We will now find the distance of the next class’s elements from each element of list S, add them up and select the minimum distance among the classes. Since, there is only one element present in the remaining dataset, so we store 1 from class C, after calculating the distance 5+8 = 13, to list S as shown in Fig. ~\ref{fig4}. Now S has {2 from class A, 2 from class B, and 1 from class C}. We continue this process for the next class, by calculating node 1 of class D from every element of S. We calculate this distance as 6+6+3 = 15. So, the final list S have the points {2 from class A, 2 from class B, 1 from class C, 1 from class D} as shown in Fig. ~\ref{fig5}. We draw the lines to separate it from other points and visualise the cluster in a map later in this paper for real world data.

If there is no member of a class in an area, then that particular class is not considered and the data is skipped. All the classes (preferences) are treated equally in importance by the algorithm.

\subsection{Collecting data for Wisp application}

The algorithm for the JJCluster is implemented using Python3. It uses Folium ~\cite{R31} ~\cite{R32} library to visualise geospatial data. Social media is generating large amount of data, after careful analysis of the data, it can be used for mapping and understanding of the evolving human landscape ~\cite{R33}. These data provide ambient geospatial information such as momentary social hotspots according to majority preferences. These provide a better understanding of those features of a place, which cannot be estimated easily.

\begin{figure}[tph!]
	
	\begin{center}                         
		\includegraphics[totalheight=8cm]{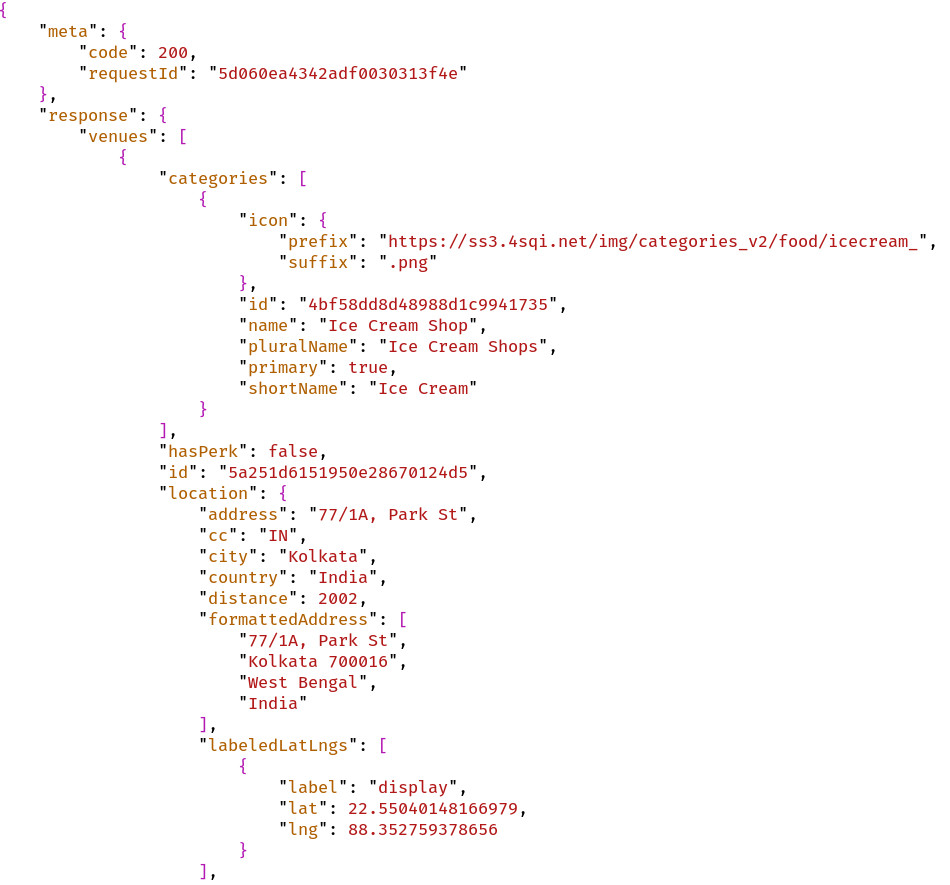}
		\caption{Restful JSON dumps of foursquare API for pre-processing}
		
		\label{fig6}
		
	\end{center}
\end{figure}

The data is collected through Foursquare ~\cite{R34}~\cite{R35} API, which is a social location service allowing users to explore world around them. It is like a general social media platform, where users can connect to their friends having foursquare account by downloading the app in mobile devices. The foursquare API has restful services, which helps the developers to create certain applications based on the data as shown in Fig. ~\ref{fig6}. The availability of foursquare is 24 x 7, and it is maintained by rich community of users. There are 150,000 developers currently building location-aware experiences with foursquare technology and data ~\cite{R29}. We have chosen foursquare API over other API in this application because we can focus on a certain location and get the data in brief.

\begin{figure}[tph!]
	
	\centering
	\includegraphics[totalheight=3cm,width=8.5cm]{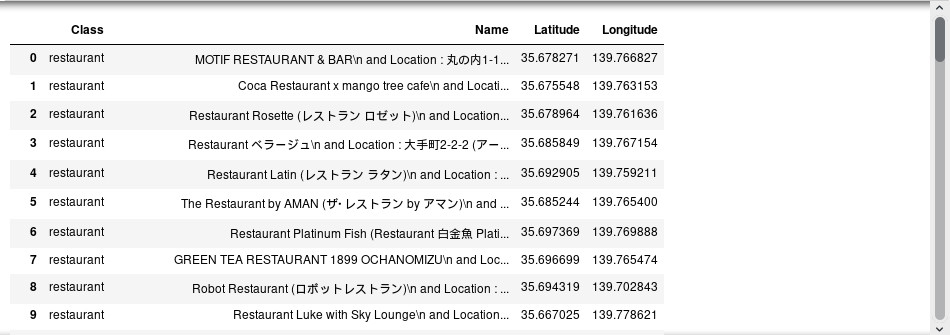}
	\caption{The formatted structured data which shall be passed as input to the JJCluster module}
	
	\label{fig7}

\end{figure}

The cleaned JSON dump is shown in Fig. ~\ref{fig7}. This data is an instance of a panda’s data frame, fed to the JJCluster module for further processing and visualisation. This is an instance of pre-processed data for the location Tokyo. The data is divided into class, name, latitude and longitude. This forms the raw data for generic clustering module. Any data following this structure can be passed as input to the JJCluster module.

\begin{figure}[tph!]
	
	\begin{center}   
		\includegraphics[totalheight=6cm]{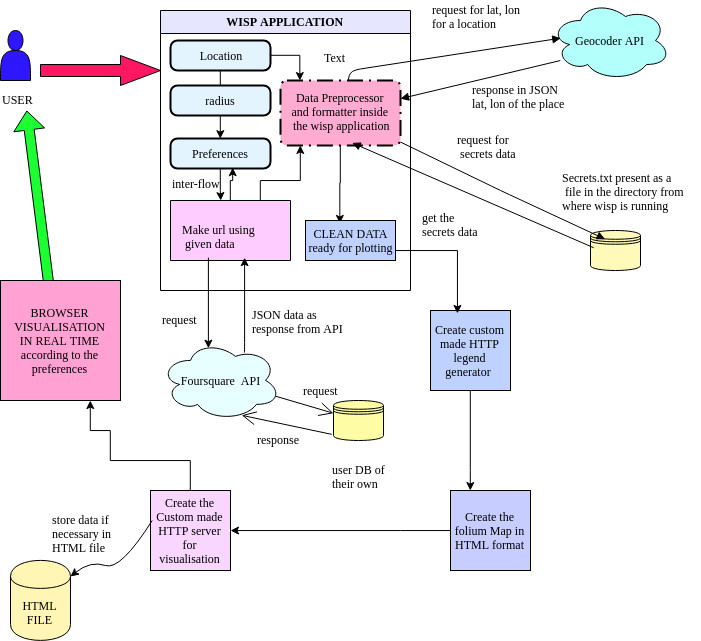}
		\caption{The internal structure of Wisp application}
		
		\label{fig8}
		
	\end{center}   
\end{figure}

The user interface here is self-intuitive, passing the necessary data as inputs to the console for visualisation. The user gives the radius, location, preferences as input as shown in Fig. ~\ref{fig8}. The location’s latitude and longitude is looked up from a remote server using Geocoder API. The foursquare credentials are taken from an external file and the preferences are requested one by one. The data is received in JSON dumps, cleaned and are formed in a formatted preference tree as input to the JJCluster module. The distance is then optimised using the algorithm; the folium map is generated along with the preference legends. The real-time map is generated and visualised using a custom made HTTP server. This map can be saved for later reference. The whole working process of the Wisp application is shown in Fig. ~\ref{fig8}. It is a stand-alone semi web based loosely coupled application, which can be integrated with a web based framework for serving a huge number of audiences in the future.

After successful fetching and cleaning the data, with the help of foursquare credentials, we use the JJCluster to find the cluster. The algorithm for the JJCluster is discussed below.

\subsection{Pseudo-code for the JJCluster algorithm in Wisp}

The algorithm for JJCluster module is written in Algorithm 1.

\begin{algorithm}
	\SetAlgoLined
	
	\KwResult{Returns a list of unique nodes from each class }
	
	\textbf{full\_tree} = [
	$[class1, [[ name11, lat11, lon11], [name12, lat12, lon12],$
	$[name13, lat13, lon13], ...]$,
	$[class2, [[ name21, lat21, lon21], [name22, lat22, lon22], $
	$ [name23, lat23, lon23], ...]$,
	$[class3, [[ name31, lat31, lon31], [name32, lat32, lon32], $
	$[name33, lat33, lon33], ...]$,
	$...$
	]\;	
	\textbf{S} = $\phi$\;
	
	\For{ \textbf{class\_list} in \textbf{full\_tree} }{
		\textbf{distance\_list} = $\phi$ \;
		\textbf{this\_class} = select class from \textbf{class\_list}\;
		\eIf{\textbf{first iteration}}{
			\For{\textbf{item} in \textbf{full\_tree}}{
				\textbf{item\_class} = select class from \textbf{item}\;
				\If{ \textbf{item\_class} $\neq$ \textbf{this\_class}}{
					\textbf{distance} = \textbf{harvesine$($} coordinates of \textbf{item}, coordinates of \textbf{item} in \textbf{class\_list} \textbf{$)$}\;
					\textbf{distance\_list} = \textbf{distance\_list} $\cup$ \textbf{distance}\;
				}
			}
		}{
			\For{\textbf{item} in \textbf{this\_class}}{
				\textbf{distance} = \textbf{harvesine$($} coordinates of \textbf{item}, coordinates of \textbf{item} in \textbf{S} \textbf{$)$}\;
				\textbf{distance\_list} = \textbf{distance\_list} $\cup$ \textbf{distance}\;
			}
		}
		\textbf{s} = find minimum \textbf{distance} from \textbf{distance\_list}\;
		\textbf{S} = \textbf{S} $\cup$ \textbf{s}\;
	}
	\textbf{S} = contains all the unique nodes from each class of the list \textbf{full\_tree}\;
	\caption{JJCluster}
	
\end{algorithm}

Here, the great circle distance between two latitude and longitude is calculated with the help of harvesine ~\cite{R35} formula, which gives accurate distance between two coordinates in a map. This algorithm results in successful formation of the best preference cluster and visualises it in a folium map. This is the general cartographic algorithm for wisp. This algorithm can be used in different fields of scientific studies for finding out the optimal points by selecting the unique nodes of each of the unique classes. We will now discuss the visualisations generated using Wisp application.

\section{Results and Discussions}

\subsection{Visualisations generated in Wisp}

\begin{figure}
	\begin{subfigure}{.5\textwidth}
		\includegraphics[totalheight=12.5cm, width=7.5cm]{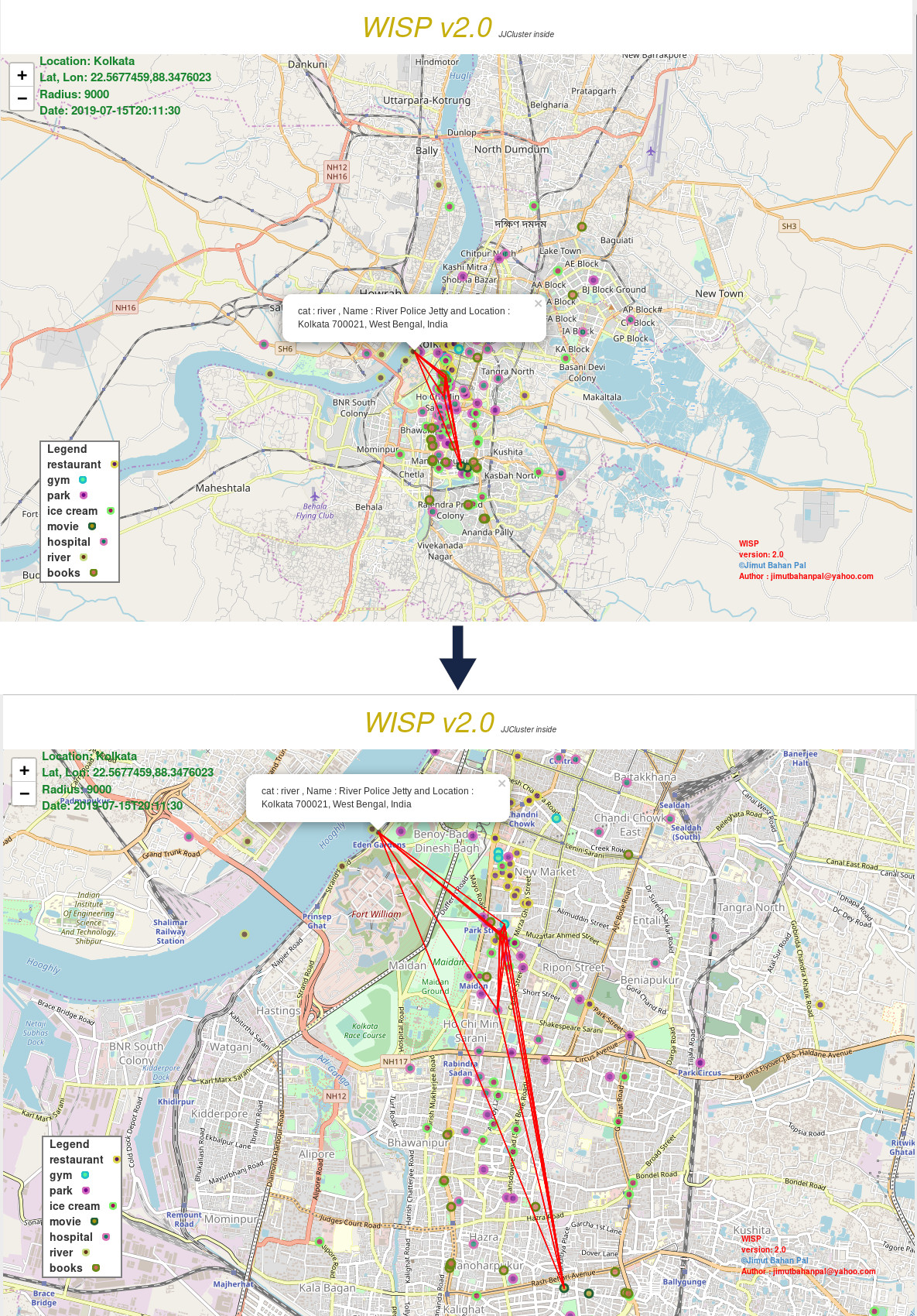}
		\caption{Open Street map of \href{https://jimut123.github.io/blogs/JJC_WISP/Tokyo_15.html}{Tokyo} with 15 preferences in a radius of 9 Km (above: Zoomed out, below: Zoomed in).}
		\label{fig9}
	\end{subfigure}%
	\begin{subfigure}{.5\textwidth}
		\includegraphics[totalheight=12.5cm, width=7.5cm]{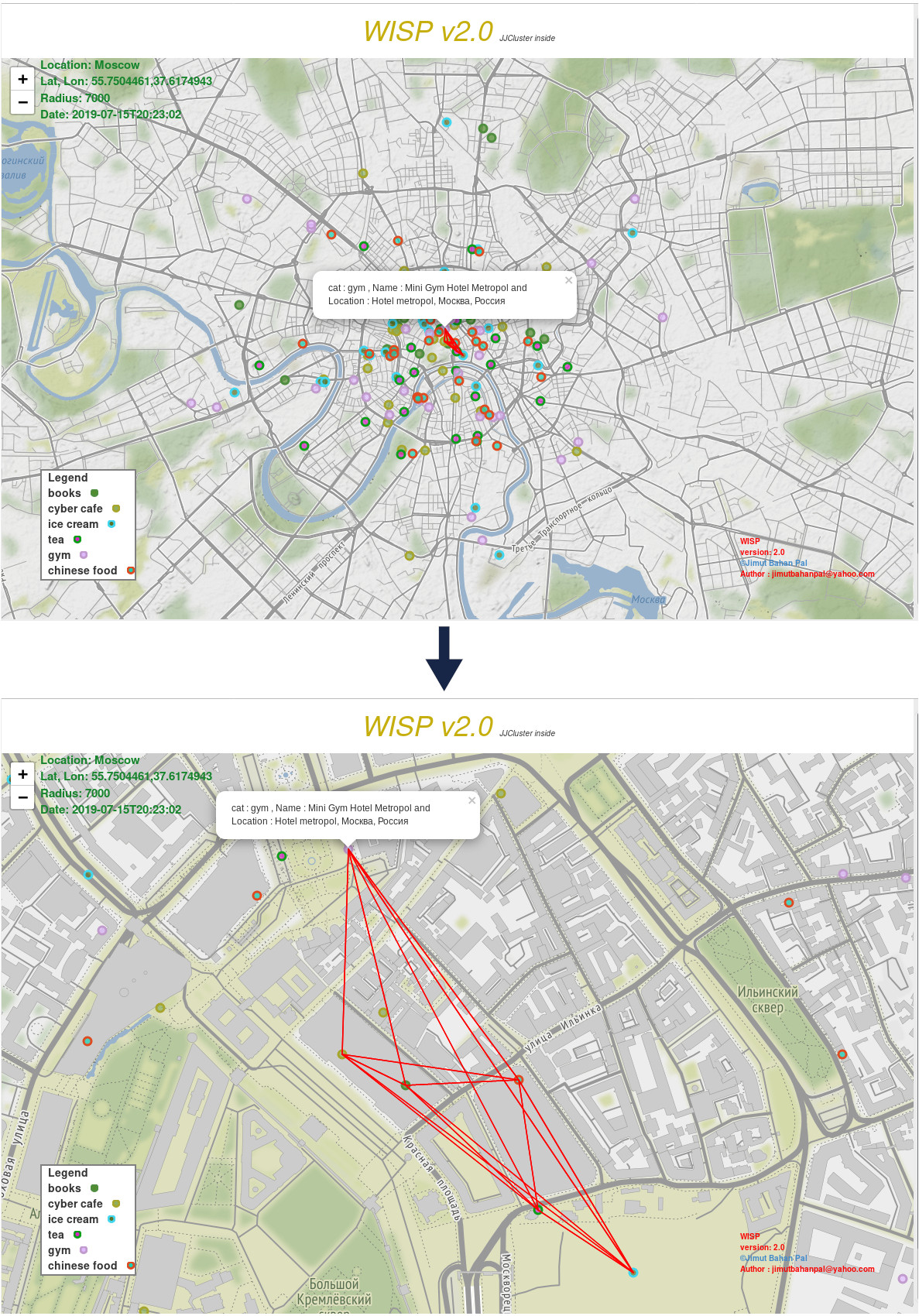}
		\caption{Open Street map of \href{https://jimut123.github.io/blogs/JJC_WISP/Kolkata_8.html}{Kolkata} with 8 preferences in a radius of 9 Km (above: Zoomed out, below: Zoomed in)}
		
		\label{fig10}
	\end{subfigure}

	\caption{A figure containing the maps formed by the Wisp application for Tokyo (left) and Kolkata (right).}
	\label{fig:maps}
\end{figure}

\begin{figure}[h!]
	
	\begin{subfigure}[t]{.5\textwidth}
		\includegraphics[totalheight=12.5cm, width=7.5cm]{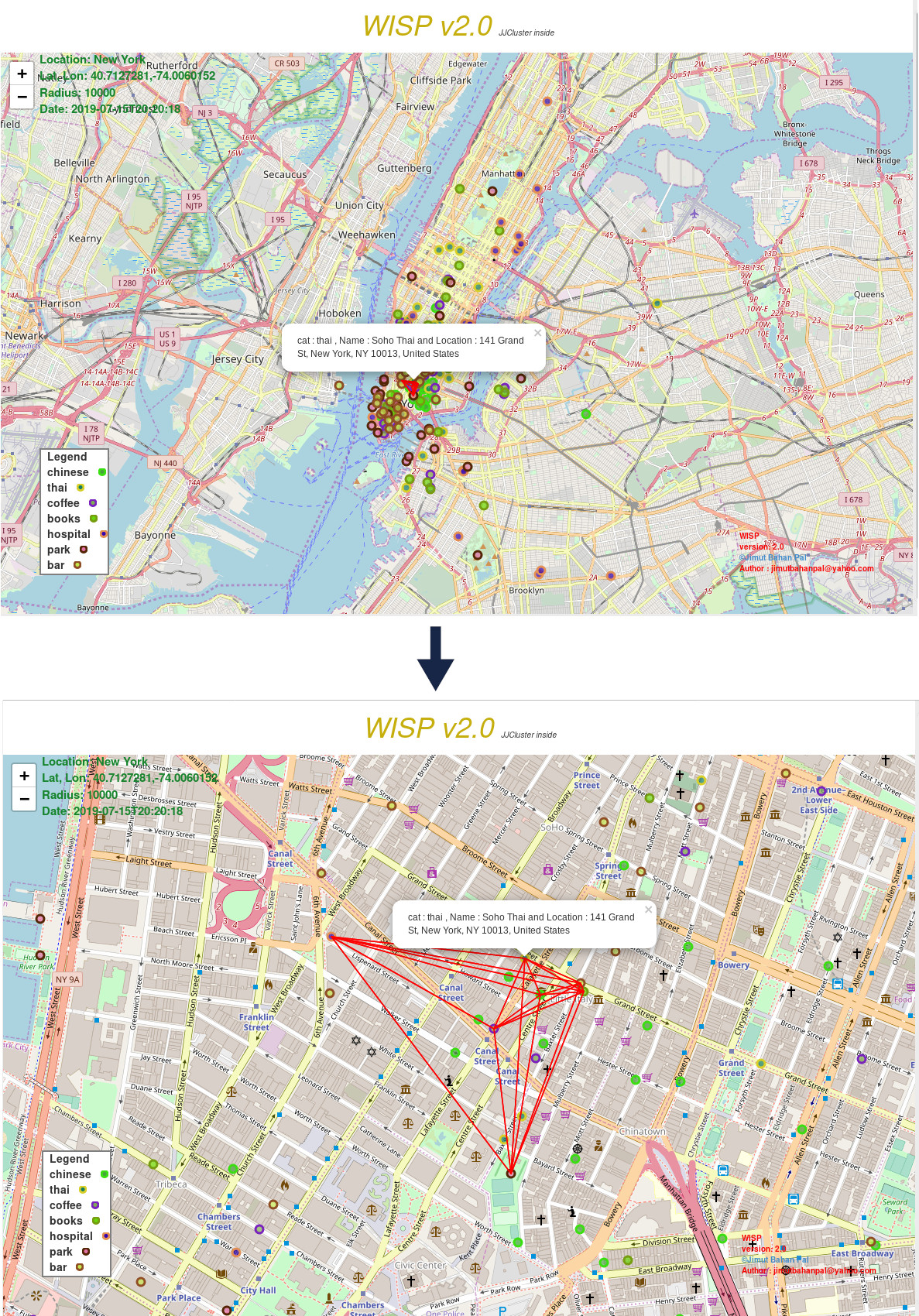}
		\caption{Stamen Terrain map of \href{https://jimut123.github.io/blogs/JJC_WISP/Moscow_6.html}{Moscow} with 6 preferences in a radius of 7 Km (above: Zoomed out, below: Zoomed in).}
		
		\label{fig11}
	\end{subfigure}\hfill
	\begin{subfigure}[t]{.5\textwidth}
		\includegraphics[totalheight=12.5cm, width=7.5cm]{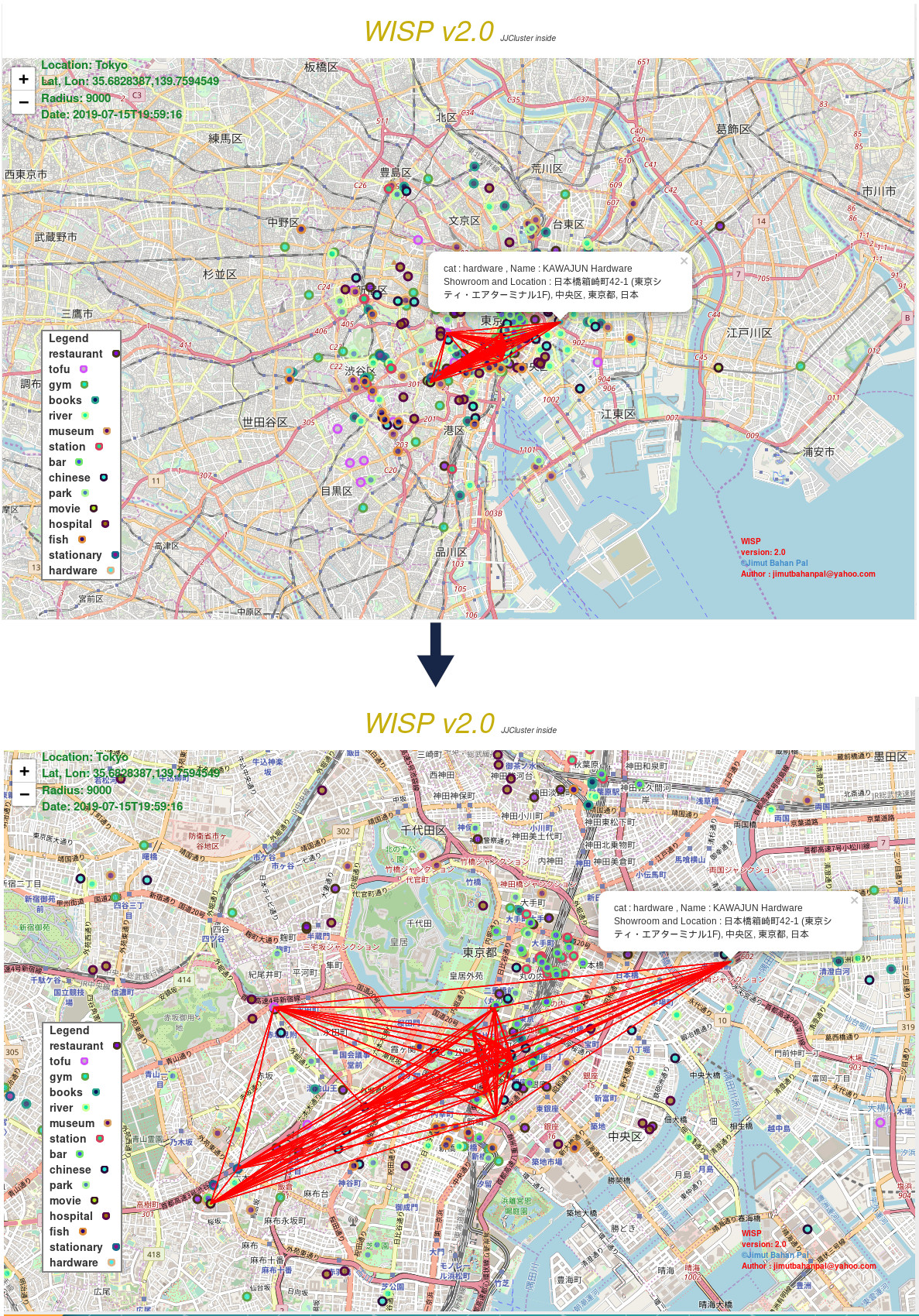}
		\caption{Open Street map of \href{https://jimut123.github.io/blogs/JJC_WISP/New_York_7.html}{New York} with 7 preferences in a radius of 10 Km (above: Zoomed out, below: Zoomed in).}
		
		\label{fig12}
	\end{subfigure}
	
	\caption{A figure containing the maps formed by the Wisp application for Moscow (left) and New York (right).}
	\label{fig:maps1}
\end{figure}

Wisp is a small application of the JJCluster algorithm. It takes foursquare secret, foursquare ID, name of the location, number of preferences, preference list, and the type of map as input and displays the map by a custom made HTTP server on real time as shown in Fig. ~\ref{fig6}. The source code of the Wisp application is uploaded in GitHub. The quality and correctness of the visualisation will depend on the data. Since, foursquare is community maintained and trusted by developers, it should provide accurate information about a place. There are certain places which are not listed by the foursquare API, or the number of reviews is too low. Such places are usually under developed and such organization’s owner probably doesn’t list themselves in the foursquare API platform. The porting of this algorithm with Google maps data may lead to more accurate results, the Google maps enlists almost all the places present in any given location. Since there is an easy premium integration of Google services, this clustering algorithm could be used by Google map data to give highly accurate map results. The wisp application is free alongside with the regular data that is fetched from foursquare API. There is a premium package available in foursquare and might cost money, but using premium Google services to get pin point location of places will be a better option for getting accurate results from this clustering algorithm. The maps that are obtained by using foursquare free services and are implemented using Python3 are shown in Fig. ~\ref{fig9}, Fig. ~\ref{fig10}, Fig. ~\ref{fig11}, and Fig. ~\ref{fig12}. The preferences are plotted in a legend for each map on the left side, which forms a unique class. Each of the data points in the map are treated as individual nodes in JJCluster. The minimum nodes are selected and are successfully plotted in the folium map of the countries.

One of the many visualisation generated by Wisp application is shown in Fig. ~\ref{fig9}, targeting Tokyo, Japan with 15 preferences within a radius of 9 km, which forms the biggest size polygon among all the maps. Since, the number of preferences are high, the zonal boundary of the area increases because all the places may not be present in a small span of area. We can see the map of Kolkata, India in Fig. ~\ref{fig10}. There are eight preferences i.e., restaurant, gym, park, ice cream, movie theatre, hospital, river and books that can be treated as unique set of classes in a map. This is visualised in a browser. We can see details of each data points with the help of pop-ups. The total amount of data points shows the popularity of the foursquare location in that place. The whole map is in a radius of 10 km, but a certain part of the map is shown. We can also determine the high market value of property such as houses and flats in such places if it is close to many preferences.Similarly, in Fig. ~\ref{fig11}, we see the small part of Russia that is clustered and zoned (among all the bigger part of the data points) according to the algorithm. We studied a more diverse domain of classes in Fig. ~\ref{fig11}, for New York. We may find different shapes and sizes of zonal polygons in a city for different preferences. It may even happen that the data are updated in foursquare API and the shapes of the polygon changes over time. Since this is a community maintained API, it helps users to get the most recent data according to their needs.

\subsection{Optimisation matrix of the visualisations}

The names of the places are treated as unique nodes, minimum distance is found by selecting the nodes from different classes and adding them to the selected class’s list S. We optimised the distance by applying the JJCluster algorithm. We define certain terms related to optimization matrix. 

\textbf{D}: The distance of the considered node to all the nodes in the list S.

\textbf{T}: The total distance of all the nodes in a particular class to all the nodes in the list S.

\textbf{k}: $\frac{T}{D}$, the optimization factor.

The optimisation matrix for Tokyo is shown in Table 1. Here we name the list S by its class as there is a constraint of place in the table. We note that it contains the name of those points, which are represented aside for its corresponding class. As per the algorithm, the list S takes the minimum of the selected points and hence, the size of list increases after each iteration.



\begin{table*}[]
	
	\centering
	
	\resizebox{\textwidth}{!}{%
		
		\begin{tabular}{|c|c|c|c|c|c|c|c|c|}
			
			\hline
			
			\textbf{Sl. No.} & \textbf{Name} & \textbf{Class} & \textbf{List S} & \textbf{Latitude} & \textbf{Longitude} & \textbf{Distance (D)} & \textbf{Total Distance (T)} & \textbf{Optimization Factor (k)} \\ \hline
			
			1 & restaurant prunier & restaurant & {[}restaurant{]} & 35.677701 & 139.761121 & 1058.644523 & 39840.316139 & 37.633328 \\ \hline
			
			2 & Chen Mapo Tofu & tofu & {[}restaurant, tofu{]} & 35.677815 & 139.736694 & 2.207082 & 68.494887 & 31.034142 \\ \hline
			
			3 & Gym & gym & \begin{tabular}[c]{@{}c@{}}{[}restaurant, tofu,\\ gym{]}\end{tabular} & 35.675640 & 139.758585 & 2.316799 & 266.431336 & 114.999777 \\ \hline
			
			4 & HMV and Books & books & \begin{tabular}[c]{@{}c@{}}{[}restaurant, tofu,\\ gym, books{]}\end{tabular} & 35.673216 & 139.759761 & 2.949521 & 378.860547 & 128.448172 \\ \hline
			
			5 & river friends & river & \begin{tabular}[c]{@{}c@{}}{[}restaurant, tofu,\\ gym, books, river{]}\end{tabular} & 35.667960 & 139.761283 & 5.050077 & 448.085848 & 88.728517 \\ \hline
			
			6 & Jansem Museum & museum & \begin{tabular}[c]{@{}c@{}}{[}restaurant, tofu,\\ gym, books, river,\\ museum{]}\end{tabular} & 35.671814 & 139.761493 & 4.143950 & 616.359580 & 148.737205 \\ \hline
			
			7 & Tabitus Station & station & \begin{tabular}[c]{@{}c@{}}{[}restaurant, tofu,\\ gym, books, river,\\ museum, station{]}\end{tabular} & 35.674524 & 139.761528 & 4.170173 & 457.917897 & 109.807887 \\ \hline
			
			8 & Peter: The Bar & bar & \begin{tabular}[c]{@{}c@{}}{[}restaurant, tofu,\\ gym, books, river,\\ museum, station, bar{]}\end{tabular} & 35.674652 & 139.760642 & 4.081305 & 289.851438 & 71.019295 \\ \hline
			
			9 & \begin{tabular}[c]{@{}c@{}}Kozanro Chinese \\ restaurant\end{tabular} & chinese & \begin{tabular}[c]{@{}c@{}}{[}restaurant, tofu,\\ gym, books, river,\\ museum, station, bar,\\ chinese{]}\end{tabular} & 35.673575 & 139.762887 & 4.907414 & 726.208372 & 147.981895 \\ \hline
			
			10 & \begin{tabular}[c]{@{}c@{}}Tokyo FM Ginza \\ Park\end{tabular} & park & \begin{tabular}[c]{@{}c@{}}{[}restaurant, tofu,\\ gym, books, river,\\ museum, station, bar,\\ chinese, park{]}\end{tabular} & 35.672573 & 139.763209 & 5.351536 & 813.479012 & 152.008497 \\ \hline
			
			11 & TOHO Cinemas & movie & \begin{tabular}[c]{@{}c@{}}{[}restaurant, tofu,\\ gym, books, river,\\ museum, station, bar,\\ chinese, park, movie{]}\end{tabular} & 35.660054 & 139.729657 & 31.137526 & 120.719115 & 3.876965 \\ \hline
			
			12 & \begin{tabular}[c]{@{}c@{}}Toranomon \\ Hospital\end{tabular} & hospital & \begin{tabular}[c]{@{}c@{}}{[}restaurant, tofu,\\ gym, books, river,\\ museum, station, bar,\\ chinese, park, movie, hospital{]}\end{tabular} & 35.668780 & 139.746678 & 16.129829 & 1061.389711 & 66.802913 \\ \hline
			
			13 & Rock Fish & fish & \begin{tabular}[c]{@{}c@{}}{[}restaurant, tofu,\\ gym, books, river,\\ museum, station, bar,\\ chinese, park, movie, hospital,\\ fish{]}\end{tabular} & 35.670040 & 139.759960 & 10.701534 & 1154.081905 & 107.842662 \\ \hline
			
			14 & \begin{tabular}[c]{@{}c@{}}Alpha note \\ stationary\end{tabular} & stationary & \begin{tabular}[c]{@{}c@{}}{[}restaurant, tofu,\\ gym, books, river,\\ museum, station, bar,\\ chinese, park, movie, hospital,\\ fish, stationary {]}\end{tabular} & 35.670040 & 139.759960 & 31.636799 & 180.066801 & 5.691688 \\ \hline
			
			15 & \begin{tabular}[c]{@{}c@{}}Kawajun Hardware\\ Showroom\end{tabular} & hardware & \begin{tabular}[c]{@{}c@{}}{[}restaurant, tofu,\\ gym, books, river,\\ museum, station, bar,\\ chinese, park, movie, hospital,\\ fish, stationary, hardware {]}\end{tabular} & 35.681928 & 139.787323 & 45.477805 & 156.733348 & 3.446370 \\ \hline
			
		\end{tabular}%
		
	}
	
	\caption{Optimisation matrix for Tokyo using JJCluster}
\end{table*}




\begin{table*}[!tph]
	
	\centering

	\resizebox{\textwidth}{!}{%
		
		\begin{tabular}{|c|c|c|c|c|c|c|c|c|}
			
			\hline
			
			\textbf{Sl. No.} & \textbf{Name} & \textbf{Class} & \textbf{List S} & \textbf{Latitude} & \textbf{Longitude} & \textbf{Distance (D)} & \textbf{Total Distance (T)} & \textbf{Optimization Factor (k)} \\ \hline
			
			1 & \begin{tabular}[c]{@{}c@{}}Oasis Restaurant,\\ Park Street\end{tabular} & restaurant & {[}restaurant{]} & 22.553118 & 88.352491 & 390.372423 & 14127.761798 & 36.181202 \\ \hline
			
			2 & \begin{tabular}[c]{@{}c@{}}Aura Gym,\\ Park Street\end{tabular} & gym & {[}restaurant, gym{]} & 22.554730 & 88.352216 & 0.181536 & 3.617280 & 19.925980 \\ \hline
			
			3 & \begin{tabular}[c]{@{}c@{}}Elliot Park ,\\ Park Street\end{tabular} & park & {[}restaurant, gym, park{]} & 22.553883 & 88.352672 & 0.229147 & 121.066560 & 528.335549 \\ \hline
			
			4 & \begin{tabular}[c]{@{}c@{}}Metro Ice Cream,\\ Park Street\end{tabular} & ice cream & \begin{tabular}[c]{@{}c@{}}{[}restaurant, gym, park,\\  ice cream{]}\end{tabular} & 22.553568 & 88.352151 & 0.250870 & 304.604499 & 1214.192773 \\ \hline
			
			5 & \begin{tabular}[c]{@{}c@{}}UFO Moviez India\\ Ltd., 68, Purna Das \\ Rd, Triangular Park\end{tabular} & movie & \begin{tabular}[c]{@{}c@{}}{[}restaurant, gym, park,\\  ice cream, movie{]}\end{tabular} & 22.517512 & 88.358810 & 16.387831 & 33.253821 & 2.029178 \\ \hline
			
			6 & \begin{tabular}[c]{@{}c@{}}Nightangle Hospital,\\ Shakespeare Sarani\end{tabular} & hospital & \begin{tabular}[c]{@{}c@{}}{[}restaurant, gym, park,\\  ice cream, movie, \\ hospital{]}\end{tabular} & 22.545964 & 88.351471 & 6.763714 & 545.871007 & 80.705805 \\ \hline
			
			7 & River Ploice Jetty & river & \begin{tabular}[c]{@{}c@{}}{[}restaurant, gym, park,\\  ice cream, movie, \\ hospital, river{]}\end{tabular} & 22.564127 & 88.338234 & 15.359777 & 196.605708 & 12.800037 \\ \hline
			
			8 & \begin{tabular}[c]{@{}c@{}}Oxford Bookstore,\\ Park Street\end{tabular} & books & \begin{tabular}[c]{@{}c@{}}{[}restaurant, gym, park,\\  ice cream, movie, \\ hospital, river,books{]}\end{tabular} & 22.553652 & 88.351732 & 7.049999 & 653.233020 & 92.657174 \\ \hline

		\end{tabular}%
		
	}
	
	\caption{Optimisation matrix for Kolkata using JJCluster}
\end{table*}



\begin{table*}[]
	
	\centering
	
	\resizebox{\textwidth}{!}{%
		
		\begin{tabular}{|c|c|c|c|c|c|c|c|c|}
			
			\hline
			
			\textbf{Sl. No.} & \textbf{Name} & \textbf{Class} & \textbf{List S} & \textbf{Latitude} & \textbf{Longitude} & \textbf{Distance (D)} & \textbf{Total Distance (T)} & \textbf{Optimization Factor (k)} \\ \hline
			
			1 & Coffee and Books & books & {[}books{]} & 55.754403 & 37.622445 & 354.560790 & 10618.142223 & 29.947311 \\ \hline
			
			2 & Bosco Cafe & cyber cafe & {[}books,cyber cafe{]} & 55.754878 & 37.620674 & 0.122773 & 61.572129 & 501.513267 \\ \hline
			
			3 & Ice Cave & ice cream & \begin{tabular}[c]{@{}c@{}}{[}books,cyber cafe,\\ ice cream{]}\end{tabular} & 55.751477 & 37.628756 & 1.143593 & 146.870961 & 128.429429 \\ \hline
			
			4 & Moscow Tea & tea & \begin{tabular}[c]{@{}c@{}}{[}books,cyber cafe,\\ ice cream, tea{]}\end{tabular} & 55.752457 & 37.626123 & 0.948188 & 195.201380 & 205.867904 \\ \hline
			
			5 & \begin{tabular}[c]{@{}c@{}}Mini Gym Hotel \\ Metropol\end{tabular} & gym & \begin{tabular}[c]{@{}c@{}}{[}books,cyber cafe,\\ ice cream, tea, gym{]}\end{tabular} & 55.758079 & 37.620857 & 2.369243 & 393.457266 & 166.068773 \\ \hline
			
			6 & \begin{tabular}[c]{@{}c@{}}Royal Chinese \\ Restaurant\end{tabular} & \begin{tabular}[c]{@{}c@{}}chinese \\ food\end{tabular} & \begin{tabular}[c]{@{}c@{}}{[}books,cyber cafe,\\ ice cream, tea, gym,\\ chinese food{]}\end{tabular} & 55.754480 & 37.625580 & 1.620614 & 284.816591 & 175.746076 \\ \hline
			
		\end{tabular}%
		
	}
	
	\caption{Optimisation matrix obtained for Moscow using JJCluster}
\end{table*}



\begin{table*}[]
	
	\centering
	
	\resizebox{\textwidth}{!}{%
		
		\begin{tabular}{|c|c|c|c|c|c|c|c|c|}
			
			\hline
			
			\textbf{Sl. No.} & \textbf{Name} & \textbf{Class} & \textbf{List S} & \textbf{Latitude} & \textbf{Longitude} & \textbf{Distance (D)} & \textbf{Total Distance (T)} & \textbf{Optimization Factor (k)} \\ \hline
			
			1 & \begin{tabular}[c]{@{}c@{}}Museum of Chinese\\ in America\end{tabular} & chinese & {[}chinese{]} & 40.719361 & -73.999086 & 448.537714 & 15335.590138 & 34.190191 \\ \hline
			
			2 & Soho Thai & thai & {[}chinese, thai{]} & 40.720117 & -73.999504 & 0.091115 & 70.532027 & 774.101656 \\ \hline
			
			3 & Kaigo Coffee Room & coffee & \begin{tabular}[c]{@{}c@{}}{[}chinese, thai, \\ coffee{]}\end{tabular} & 40.718633 & -74.000367 & 0.315488 & 105.182689 & 333.396911 \\ \hline
			
			4 & \begin{tabular}[c]{@{}c@{}}Indigo Books HQ\\ Design Studio\end{tabular} & books & \begin{tabular}[c]{@{}c@{}}{[}chinese, thai, \\ coffee, books{]}\end{tabular} & 40.719422 & -73.997885 & 0.485303 & 223.256302 & 460.035156 \\ \hline
			
			5 & \begin{tabular}[c]{@{}c@{}}Tribeca Soho \\ Animal Hospital\end{tabular} & hospital & \begin{tabular}[c]{@{}c@{}}{[}chinese, thai, \\ coffee, books, hospital{]}\end{tabular} & 40.720578 & -74.004869 & 2.000487 & 586.780255 & 293.318670 \\ \hline
			
			6 & Columbus Park & park & \begin{tabular}[c]{@{}c@{}}{[}chinese, thai, \\ coffee, books, hospital,\\ park{]}\end{tabular} & 40.715603 & -73.999884 & 2.417840 & 360.109107 & 148.938377 \\ \hline
			
			7 & \begin{tabular}[c]{@{}c@{}}Onieal's Grand Street\\ Bar and Restaurant\end{tabular} & bar & \begin{tabular}[c]{@{}c@{}}{[}chinese, thai, \\ coffee, books, hospital,\\ park, bar{]}\end{tabular} & 40.719593 & -73.997966 & 1.553863 & 196.666538 & 126.566175 \\ \hline
			
		\end{tabular}%
		
	}
	
	\caption{Optimisation matrix for New York using JJCluster}
\end{table*}

The optimisation matrix corresponds to actual data fetched from the foursquare API, and scaled in the map accordingly. We see the selected node’s list increase every time we increase the iteration. Similarly, Table 2, 3 and 4 contains the visualizations obtained from the cities Kolkata, Moscow, and New York respectively. The minimum distance may likewise increase or decrease depending upon the total closeness of a point to that list. This algorithm has a complexity of $\theta$($n^{2}$), we may improve this algorithm in future. These visualisations can be viewed in web browser and in smart phone. Our next target is to build an application for serving in the web, or maybe alongside with mobile applications.

The main objective for designing this Wisp application is to eliminate incorrect manual search. This clustering algorithm merely needs data in a particular format, so hacking and tweaking it for suitable uses are encouraged, since this is an open sourced project.

\section{Conclusions and Future Scope}

Creation of real time visualisation application is successful using JJCluster algorithm. We studied an application domain of the JJCluster algorithm for selecting the best places according to preferences in a map using Wisp application. The main objective of designing this JJCluster algorithm is to optimise and select the data points in such a way that it minimises the total distance of every nearest nodes from different classes. This algorithm can be used by researchers of different fields like statisticians, economists, cartographers, general scientists who are interested in finding certain hotspots according to classes. This may also be used to deploy drone swarms in those locations, where critical action is needed, which is determined by the algorithm, since this minimizes the total distance. Applications for this algorithm may further attract interdisciplinary researchers to evolve this algorithm according to their needs. It may happen that an oceanographer is studying about a place and there are too many classes of objects in his research. The objects with unique classes may be scattered over a wide place, but using this algorithm he may be able to select and find those places, which have all of the unique classes in the shortest span of area of the map. This will help to reduce fuel, time, resources and focus on areas of concern, having all the resources, which he might want to study.

Wisp is built on a map based system, it is open sourced, and scientists can use this module by passing necessary formatted data to give visualisations in real time. The real fuel here is data, if the data is accurate, it will give good results, but it may happen that certain places doesn’t have the data, in that case use of premium Google services may dominate than using foursquare free services for wisp application. The visualisation is done in real time and saves users from manually searching a place. Since foursquare is community maintained API, it may happen that certain labelling of data is wrong, as the users have freedom to add and manipulate data. We may use Google API’s in future for its popularity and correctness. In conclusion, we just need to pass the necessary formatted data to get the visualisations in real time with the help of Wisp application. In this method we can cluster nearest unique nodes from different classes using JJCluster algorithm in Wisp application.

\section{Acknowledgements}

I am thankful to my father, Dr. Jadab Kumar Pal, Deputy Chief Executive, Indian Statistical Institute, Kolkata, for constant motivation and support for completion of this project. I dedicate this clustering algorithm to my brother Jisnoo Dev Pal. I also acknowledge Prof Shalabh Agarwal, former Head of the department of Computer Science, St. Xavier's College, Kolkata for his support. Finally, I am thankful to all my well-wishers who supported me in every part of this project. The source code for Wisp using JJCluster can be found here : \href{https://github.com/Jimut123/wisp}{https://github.com/Jimut123/wisp}.

\bibliographystyle{plain}
\bibliography{main}

\end{document}